\documentclass[aps,preprint,nofootinbib,floatfix]{revtex4}%
\usepackage{amssymb}
\usepackage{amsmath}
\usepackage{epsfig}
\usepackage{graphicx}
\usepackage{hyperref}
\usepackage{amsfonts}%
\begin{document}
\title{Linear Relations of High Energy Absorption/Emission Amplitudes of D-brane}
\author{Jen-Chi Lee}
\email{jcclee@cc.nctu.edu.tw}
\affiliation{Department of Electrophysics, National Chiao-Tung University, Hsinchu, Taiwan, R.O.C.}
\author{Yi Yang}
\email{yyang@phys.cts.nthu.edu.tw}
\affiliation{Department of Electrophysics, National Chiao-Tung University and Physics
Division, National Center for Theoretical Sciences, Hsinchu, Taiwan, R.O.C.}
\date{\today }

\begin{abstract}
We calculate the absorption amplitudes of a closed string state at arbitrary
mass level leading to two open string states on the D-brane at high energies.
As in the case of Domain-wall scattering we studied previously, this process
contains only one kinematic variable. However, in contrast to the power-law
behavior of Domain-wall scattering, its form factor behaves as exponential
fall-off in the high energy limit. After identifying the geometric parameter
of the kinematic, we derive the linear relations (of the kinematic variable)
and ratios among the high energy amplitudes corresponding to absorption of
different closed string states for each fixed mass level by D-brane. This
result is consistent with the coexistence of the linear relations and
exponential fall-off behavior of high energy string/D-brane amplitudes.

\end{abstract}
\maketitle

It is well-known that there are two fundamental characteristics of high energy
string scattering amplitudes, which make them very different from those of
field theory scatterings. The first one is the softer exponential fall-off
behavior of the form factors of string scatterings in the high-energy limit
\cite{Veneziano} in contrast to the power-law behavior of point-particle field
theory scatterings. The second one is the existence of Regge-pole structure in
the high energy string scattering amplitudes \cite{Closed} due to the infinite
number of resonances in the string spectrum.

Recently high-energy, fixed angle behavior of string scattering amplitudes
\cite{GM, Gross, GrossManes} was intensively reinvestigated for massive string
states at arbitrary mass levels \cite{ChanLee1,ChanLee2,
CHL,CHLTY,PRL,paperB,susy,Closed}. An infinite number of linear relations, or
stringy symmetries, among string scattering amplitudes of different string
states were obtained. An important new ingredient of these calculations is the
zero-norm states (ZNS) \cite{ZNS1,ZNS3,ZNS2} in the old covariant first
quantized (OCFQ) string spectrum. The existence of these infinite linear
relations constitutes the third fundamental characteristics of high energy
string scatterings, which is not shared by the usual point-particle field
theory scatterings. These linear relations persist for string scattered from
generic D$p$-brane \cite{Dscatt} except D-instanton and Domain-wall. For the
scattering of D-instanton, the form factor exhibits the well-known power-law
behavior without Regge-pole structure, and thus resembles a field theory
amplitude. For the special case of Domain-wall scattering, it was discovered
\cite{Wall} recently that, in contrast to the common wisdom \cite{Klebanov},
its form factor behaves as\textit{ power-law} with Regge-pole structure. This
discovery makes Domain-wall scatterings an unique example of a hybrid of
string and field theory scatterings. Moreover, it was shown \cite{Wall} that
the linear relations break down for the Domain-wall scattering due to this
unusual power-law behavior. This result gives a strong evidence that the
existence of the infinite linear relations, or stringy symmetries, of
high-energy string scattering amplitudes is responsible for the softer,
exponential fall-off high-energy string scatterings than the power-law field
theory scatterings. It is crucial to note that there is only one kinematic
variable for the Domain-wall scatterings in contrast to two for other generic
D$p$-brane scatterings with $p\geq0$. This is one of the main reasons that
force the high energy behavior of Domain-wall scattering to be the unusual
power-law one.

In this letter, we calculate the absorption amplitudes of a closed string
state at arbitrary mass level leading to two open string states on the D-brane
at high energies. The corresponding simple case of absorption amplitude for
massless closed string state was calculated in \cite{Decay1} (The discussion
on massless string states scattered from D-brane can be found in
\cite{Klebanov,Myers}).The inverse of this process can be used to describe
Hawking radiation in the D-brane picture. As in the case of Domain-wall
scattering discussed above, this process contains \textit{one} kinematic
variable (energy E) and thus occupies an intermediate position between the
conventional three-point and four-point amplitudes. However, in contrast to
the power-law behavior of high energy Domain-wall scattering which contains
only one kinematic variable (energy E), its form factor behaves as exponential
fall-off at high energies. It is thus of interest to investigate whether the
usual linear relations of high energy amplitudes persist for this case or not.
As will be shown in this paper, after identifying the geometric parameter of
the kinematic, one can derive the linear relations (of the kinematic variable)
and ratios among the high energy amplitudes corresponding to absorption of
different closed string states for each fixed mass level by D-brane. This
result is consistent with the coexistence \cite{Wall} of the linear relations
and exponential fall-off behavior of high energy string/D-brane amplitudes.

We first briefly review the high energy scatterings of four open string
states. At a fixed mass level $M_{op}^{2}=2(n-1)$ of 26D open bosonic string
theory, it was shown that \cite{CHLTY,PRL} a four-point function is at the
leading order at high-energy limit only for states of the following form%
\begin{equation}
\left\vert n,2m,q\right\rangle \equiv(\alpha_{-1}^{T})^{n-2m-2q}(\alpha
_{-1}^{L})^{2m}(\alpha_{-2}^{L})^{q}\left\vert 0,k\right\rangle . \label{1}%
\end{equation}
where $n\geqslant2m+2q,m,q\geqslant0.$Note that, in the high energy limit, the
scattering process becomes a plane scattering. The state in Eq.(\ref{1}) is
arbitrarily chosen to be the second vertex of the four-point function. The
other three points can be any string states. We have defined the normalized
polarization vectors of the second string state to be \cite{ChanLee1,ChanLee2}%
\begin{equation}
e^{P}=\frac{1}{M_{op}}(E_{2},\mathrm{k}_{2},0)=\frac{k_{2}}{M_{op}}, \label{2}%
\end{equation}%
\begin{equation}
e^{L}=\frac{1}{M_{op}}(\mathrm{k}_{2},E_{2},0), \label{3}%
\end{equation}%
\begin{equation}
e^{T}=(0,0,1) \label{4}%
\end{equation}
in the CM frame contained in the plane of scattering. By using the decoupling
of two types of ZNS,%
\begin{align}
\text{Type I}:L_{-1}\left\vert x\right\rangle ,  &  \text{ where }%
L_{1}\left\vert x\right\rangle =L_{2}\left\vert x\right\rangle =0,\text{
}L_{0}\left\vert x\right\rangle =0\text{ and}\\
\text{Type II}:(L_{-2}+\frac{3}{2}L_{-1}^{2})\left\vert \widetilde
{x}\right\rangle ,  &  \text{ where }L_{1}\left\vert \widetilde{x}%
\right\rangle =L_{2}\left\vert \widetilde{x}\right\rangle =0,\text{ }%
(L_{0}+1)\left\vert \widetilde{x}\right\rangle =0,
\end{align}
in the high energy limit, it was shown that there exists infinite linear
relations among string scattering amplitudes \cite{CHLTY,PRL}%
\begin{equation}
\mathcal{T}^{(n,2m,q)}=\left(  -\frac{1}{M_{op}}\right)  ^{2m+q}\left(
\frac{1}{2}\right)  ^{m+q}(2m-1)!!\mathcal{T}^{(n,0,0)}. \label{5}%
\end{equation}
Moreover, these linear relations can be used to fix the ratios among high
energy scattering amplitudes of different string states at each fixed mass
level algebraically. Eq.(\ref{5}) explicitly shows that there is only one
independent high-energy scattering amplitudes at each fixed mass level.

\begin{figure}[ptb]
\label{scattering1} \setlength{\unitlength}{3pt}
\par
\begin{center}
\begin{picture}(100,60)(-50,-30)
{
\put(-70,-20){\line(1,0){100}} \put(-30,20){\line(1,0){100}}
\put(-70,-20){\line(1,1){40}} \put(30,-20){\line(1,1){40}}
\put(-60,-18){D-brane}
\put(-40,0){\vector(1,0){80}} \put(41,0){$x$}
\put(0,0){\vector(0,1){40}} \put(0,42){$y$}
\put(18,18){\vector(-1,-1){35}} \put(-19,-19){$z$}
\put(-24,32){\vector(3,-4){24}} \put(-18,15){$k_c$}
\put(33,11){\vector(-3,-1){33}} \put(18,9){$k_2$}
\put(24,-12){\vector(-2,1){24}} \put(10,-10){$k_1$}
\put(-18,24){\vector(4,3){10}} \put(-13,31){$e^T$}
\qbezier(-5,0)(-5,3)(-3,4) \qbezier(4,-2)(7,1)(6,2)
\put(-8,2){$\phi$} \put(11,1){$\theta$} \put(9,-3){$\theta$}
\put(-25,-30){Fig.1 Kinematic setting up} }
\end{picture}
\end{center}
\end{figure}

To study the high energy process of D$p$ brane $\left(  2\leq p\leq24\right)
$ absorbs (emits) a massive closed string state leading to two open strings on
the D$p$ brane, we set up the kinematic for the massive closed string state to
be%
\begin{align}
e^{P} &  =\frac{1}{M}\left(  E,\mathrm{k}_{c}\cos\phi,-\mathrm{k}_{c}\sin
\phi,0\right)  =\frac{k_{c}}{M},\nonumber\\
e^{L} &  =\frac{1}{M}\left(  \mathrm{k}_{c},E\cos\phi,-E\sin\phi,0\right)
,\nonumber\\
e^{T} &  =\left(  0,\sin\phi,\cos\phi,0\right)  ,\nonumber\\
k_{c} &  =\left(  E,\mathrm{k}_{c}\cos\phi,-\mathrm{k}_{c}\sin\phi,0\right)  .
\end{align}
For simplicity, we chose the open string excitation to be two tachyons with
momenta (see Fig.1)%
\begin{align}
k_{1} &  =\left(  -\frac{E}{2},-\frac{\mathrm{k}_{op}}{2}\cos\theta
,0,-\frac{\mathrm{k}_{op}}{2}\sin\theta\right)  ,\\
k_{2} &  =\left(  -\frac{E}{2},-\frac{\mathrm{k}_{op}}{2}\cos\theta
,0,+\frac{\mathrm{k}_{op}}{2}\sin\theta\right)  .
\end{align}
Our final results, however, will remain the same for arbitrary two open string
excitation at high energies. Conservation of momentum on the D-brane implies%
\begin{equation}
\underset{\left(  k_{c}\right)  _{//}}{\underbrace{\frac{1}{2}\left(
k_{c}+D\cdot k_{c}\right)  }}+k_{1}+k_{2}=0\Rightarrow\mathrm{k}_{c}\cos
\phi=\mathrm{k}_{op}\cos\theta,\label{momen}%
\end{equation}
where $D_{\mu\nu}=$diag$\left\{  -1,1,-1,1\right\}  $. It is crucial to note
that, in the high energy limit, $\mathrm{k}_{c}=\mathrm{k}_{op}$ and the
scattering angle $\theta$ is identical to the incident angle $\phi$. One can
calculate$\theta$%
\begin{align}
e^{T}\cdot k_{1} &  =e^{T}\cdot k_{2}=e^{T}\cdot D\cdot k_{1}=e^{T}\cdot
D\cdot k_{2}\nonumber\\
&  =-\frac{\mathrm{k}_{op}\cos\theta\sin\phi}{2}=-\frac{\mathrm{k}_{c}\sin
\phi\cos\phi}{2},\nonumber\\
e^{L}\cdot k_{1} &  =e^{L}\cdot k_{2}=e^{L}\cdot D\cdot k_{1}=e^{L}\cdot
D\cdot k_{2}\nonumber\\
&  =\frac{1}{M}\left[  \frac{\mathrm{k}_{c}E}{2}-\frac{\mathrm{k}_{op}E}%
{2}\cos\theta\cos\phi\right]  =\frac{\mathrm{k}_{c}E}{2M}\sin^{2}%
\phi,\nonumber\\
e^{T}\cdot D\cdot k_{c} &  =2\mathrm{k}_{c}\sin\phi\cos\phi,\nonumber\\
e^{L}\cdot D\cdot k_{c} &  =-\frac{2\mathrm{k}_{c}E}{M}\sin^{2}\phi
,\label{relate}%
\end{align}
which will be useful for later calculations. We can define the kinematic
invariants%
\begin{align}
t &  \equiv-\left(  k_{1}+k_{2}\right)  ^{2}=M_{1}^{2}+M_{2}^{2}-2k_{1}\cdot
k_{2}=-2\left(  2+k_{1}\cdot k_{2}\right)  \nonumber\\
&  =2k_{1}\cdot k_{c}=2k_{2}\cdot k_{c},\\
s &  \equiv4k_{1}\cdot k_{2}=2M_{1}^{2}+2M_{2}^{2}+2\left(  k_{1}%
+k_{2}\right)  ^{2}=-2\left(  4+t\right)  ,\label{st}%
\end{align}
and calculate the following identities%
\begin{align}
k_{1}\cdot k_{c}+k_{2}\cdot D\cdot k_{c} &  =k_{2}\cdot k_{c}+k_{1}\cdot
D\cdot k_{c}=t,\\
k_{c}\cdot D\cdot k_{c} &  =M^{2}-2t.
\end{align}
Note that there is only one kinematic variable as $s$ and $t$ are related in
Eq.(\ref{st}) \cite{Decay1}. On the other hand, since the scattering angle
$\theta$ is fixed by the incident angle $\phi$, $\phi$ and $\theta$ are not
the dynamical variables in the usual sense.

Following Eq.(\ref{1}), we consider an incoming high energy massive closed
state to be \cite{Dscatt,Wall} $\left(  \alpha_{-1}^{T}\right)  ^{n-m-2q}%
\left(  \alpha_{-1}^{L}\right)  ^{m}\left(  \alpha_{-2}^{L}\right)
^{q}\otimes\left(  \tilde{\alpha}_{-1}^{T}\right)  ^{n-m^{\prime}-2q^{\prime}%
}\left(  \tilde{\alpha}_{-1}^{L}\right)  ^{m^{\prime}}\left(  \tilde{\alpha
}_{-2}^{L}\right)  ^{q^{\prime}}\left\vert 0\right\rangle $ with
\underline{$m=m^{\prime}=0$}. The amplitude of the absorption process can be
calculated to be%

\begin{align}
A  &  =\int dx_{1}dx_{2}d^{2}z\cdot\left(  x_{1}-x_{2}\right)  ^{k_{1}\cdot
k_{2}}\left(  z-\bar{z}\right)  ^{k_{c}\cdot D\cdot k_{c}}\left(
x_{1}-z\right)  ^{k_{1}\cdot k_{c}}\nonumber\\
\cdot &  \left(  x_{1}-\bar{z}\right)  ^{k_{1}\cdot D\cdot k_{c}}\left(
x_{2}-z\right)  ^{k_{2}\cdot k_{c}}\left(  x_{2}-\bar{z}\right)  ^{k_{2}\cdot
D\cdot k_{c}}\nonumber\\
&  \cdot\exp\left\{  \left\langle \left[  ik_{1}X\left(  x_{1}\right)
+ik_{2}X\left(  x_{2}\right)  +ik_{c}\tilde{X}\left(  \bar{z}\right)  \right]
\left[  \left(  n-2q\right)  \varepsilon_{T}^{\left(  1\right)  }\partial
X^{T}+iq\varepsilon_{L}^{\left(  1\right)  }\partial^{2}X^{L}\right]  \left(
z\right)  \right\rangle \right. \nonumber\\
&  +\left\langle \left[  ik_{1}X\left(  x_{1}\right)  +ik_{2}X\left(
x_{2}\right)  +ik_{c}X\left(  z\right)  \right]  \left[  \left(  n-2q^{\prime
}\right)  \varepsilon_{T}^{\left(  2\right)  }\bar{\partial}\tilde{X}%
^{T}+iq^{\prime}\varepsilon_{L}^{\left(  2\right)  }\bar{\partial}^{2}%
\tilde{X}^{L}\right]  \left(  \bar{z}\right)  \right\rangle _{\text{linear
terms}}\nonumber\\
&  =\left(  -1\right)  ^{q+q^{\prime}}\int dx_{1}dx_{2}d^{2}z\cdot\left(
x_{1}-x_{2}\right)  ^{k_{1}\cdot k_{2}}\left(  z-\bar{z}\right)  ^{k_{c}\cdot
D\cdot k_{c}}\left(  x_{1}-z\right)  ^{k_{1}\cdot k_{c}}\nonumber\\
\cdot &  \left(  x_{1}-\bar{z}\right)  ^{k_{1}\cdot D\cdot k_{c}}\left(
x_{2}-z\right)  ^{k_{2}\cdot k_{c}}\left(  x_{2}-\bar{z}\right)  ^{k_{2}\cdot
D\cdot k_{c}}\\
&  \cdot\left[  \frac{ie^{T}\cdot k_{1}}{x_{1}-z}+\frac{ie^{T}\cdot k_{2}%
}{x_{2}-z}+\frac{ie^{T}\cdot D\cdot k_{c}}{\bar{z}-z}\right]  ^{n-2q}%
\cdot\left[  \frac{ie^{T}\cdot D\cdot k_{1}}{x_{1}-\bar{z}}+\frac{ie^{T}\cdot
D\cdot k_{2}}{x_{2}-\bar{z}}+\frac{ie^{T}\cdot D\cdot k_{c}}{z-\bar{z}%
}\right]  ^{n-2q^{\prime}}\nonumber\\
&  \cdot\left[  \frac{e^{L}\cdot k_{1}}{\left(  x_{1}-z\right)  ^{2}}%
+\frac{e^{L}\cdot k_{2}}{\left(  x_{2}-z\right)  ^{2}}+\frac{e^{L}\cdot D\cdot
k_{c}}{\left(  \bar{z}-z\right)  ^{2}}\right]  ^{q}\cdot\left[  \frac
{e^{L}\cdot D\cdot k_{1}}{\left(  x_{1}-\bar{z}\right)  ^{2}}+\frac{e^{L}\cdot
D\cdot k_{2}}{\left(  x_{2}-\bar{z}\right)  ^{2}}+\frac{e^{L}\cdot D\cdot
k_{c}}{\left(  z-\bar{z}\right)  ^{2}}\right]  ^{q^{\prime}}.
\end{align}
Set $\left\{  x_{1},x_{2},z\right\}  =\left\{  -x,x,i\right\}  $ to fix the
$SL(2,R)$ gauge and use Eq.(\ref{relate}), we have%
\begin{align}
A  &  =\left(  -1\right)  ^{n+M^{2}/2+t/2}2^{M^{2}-2-5t/2}\cdot\int_{-\infty
}^{+\infty}dx\cdot x^{-t/2-2}\left(  1-ix\right)  ^{t+1}\left(  1+ix\right)
^{t+1}\nonumber\\
&  \cdot\left[  \frac{-\frac{\mathrm{k}_{c}\sin\phi\cos\phi}{2}}{1-ix}%
+\frac{-\frac{\mathrm{k}_{c}\sin\phi\cos\phi}{2}}{1+ix}+\frac{2\mathrm{k}%
_{c}\sin\phi\cos\phi}{2}\right]  ^{n-2q}\nonumber\\
&  \cdot\left[  \frac{-\frac{\mathrm{k}_{c}\sin\phi\cos\phi}{2}}{1+ix}%
+\frac{-\frac{\mathrm{k}_{c}\sin\phi\cos\phi}{2}}{1-ix}+\frac{2\mathrm{k}%
_{c}\sin\phi\cos\phi}{2}\right]  ^{n-2q^{\prime}}\\
&  \cdot\left[  \frac{\frac{\mathrm{k}_{c}E}{2M}\sin^{2}\phi}{\left(
1-ix\right)  ^{2}}+\frac{\frac{\mathrm{k}_{c}E}{2M}\sin^{2}\phi}{\left(
1+ix\right)  ^{2}}+\frac{-\frac{2\mathrm{k}_{c}E}{M}\sin^{2}\phi}{4}\right]
^{q}\cdot\left[  \frac{\frac{\mathrm{k}_{c}E}{2M}\sin^{2}\phi}{\left(
1+ix\right)  ^{2}}+\frac{\frac{\mathrm{k}_{c}E}{2M}\sin^{2}\phi}{\left(
1-ix\right)  ^{2}}+\frac{-\frac{2\mathrm{k}_{c}E}{M}\sin^{2}\phi}{4}\right]
^{q^{\prime}}\nonumber\\
&  =\left(  -1\right)  ^{n+M^{2}/2+t/2}2^{M^{2}-2-5t/2}\cdot\left(
\mathrm{k}_{c}\sin\phi\cos\phi\right)  ^{2n-2\left(  q+q^{\prime}\right)
}\left(  -\frac{\mathrm{k}_{c}E\sin^{2}\phi}{2M}\right)  ^{q+q^{\prime}%
}\nonumber\\
&  \cdot\int_{-\infty}^{+\infty}dx\cdot x^{-t/2-2}\left(  1+x^{2}\right)
^{t+1}\left[  \frac{x^{2}}{1+x^{2}}\right]  ^{2n-2\left(  q+q^{\prime}\right)
}\left[  1-\frac{2\left(  1-x^{2}\right)  }{\left(  1+x^{2}\right)  ^{2}%
}\right]  ^{q+q^{\prime}}.
\end{align}
By using the binomial expansion, we get%
\begin{align}
A  &  =\left(  -1\right)  ^{n+M^{2}/2+t/2}2^{M^{2}-2-5t/2}\cdot\left(
E\sin\phi\cos\phi\right)  ^{2n}\left(  -\frac{1}{2M\cos^{2}\phi}\right)
^{q+q^{\prime}}\nonumber\\
&  \cdot\sum_{i=0}^{q+q^{\prime}}\sum_{j=0}^{i}\binom{q+q^{\prime}}{i}%
\binom{i}{j}\left(  -2\right)  ^{i}\left(  -1\right)  ^{j}\nonumber\\
&  \int_{0}^{\infty}d\left(  x^{2}\right)  \cdot\left(  x^{2}\right)
^{-t/4-3/2+2n-2\left(  q+q^{\prime}\right)  +j}\left(  1+x^{2}\right)
^{t+1-2n+2\left(  q+q^{\prime}\right)  -2i}.
\end{align}
Finally, to reduce the integral to the standard beta function, we do the
linear fractional transformation $x^{2}=\frac{1-y}{y}$ to get
\begin{align}
&  A=\left(  -1\right)  ^{n+M^{2}/2+t/2}2^{M^{2}-2-5t/2}\cdot\left(  E\sin
\phi\cos\phi\right)  ^{2n}\left(  -\frac{1}{2M\cos^{2}\phi}\right)
^{q+q^{\prime}}\nonumber\\
&  \cdot\sum_{i=0}^{q+q^{\prime}}\sum_{j=0}^{i}\binom{q+q^{\prime}}{i}%
\binom{i}{j}\left(  -2\right)  ^{i}\left(  -1\right)  ^{j}\int_{0}^{1}dy\cdot
y^{-3t/4-3/2+2i-j}\cdot\left(  1-y\right)  ^{-t/4-3/2+2n-2\left(  q+q^{\prime
}\right)  +j}\nonumber\\
&  =\left(  -1\right)  ^{n+M^{2}/2+t/2}2^{M^{2}-2-5t/2}\cdot\left(  E\sin
\phi\cos\phi\right)  ^{2n}\left(  -\frac{1}{2M\cos^{2}\phi}\right)
^{q+q^{\prime}}\nonumber\\
&  \cdot\frac{\Gamma\left(  -\frac{3t}{4}-\frac{1}{2}\right)  \Gamma\left(
-\frac{t}{4}-\frac{1}{2}\right)  }{\Gamma\left(  -t-1\right)  }\sum
_{i=0}^{q+q^{\prime}}\sum_{j=0}^{i}\binom{q+q^{\prime}}{i}\binom{i}{j}\left(
-2\right)  ^{i}\left(  -1\right)  ^{j}\left(  \frac{3}{4}\right)
^{2i-j}\left(  \frac{1}{4}\right)  ^{2n-2\left(  q+q^{\prime}\right)
+j}\nonumber\\
&  =\left(  -1\right)  ^{n+M^{2}/2+t/2}2^{M^{2}-2-5t/2}\cdot\left(
\frac{E\sin\phi\cos\phi}{4}\right)  ^{2n}\nonumber\\
&  \cdot\left(  -\frac{2}{M\cos^{2}\phi}\right)  ^{q+q^{\prime}}\frac
{\Gamma\left(  -\frac{3t}{4}-\frac{1}{2}\right)  \Gamma\left(  -\frac{t}%
{4}-\frac{1}{2}\right)  }{\Gamma\left(  -t-1\right)  }. \label{main}%
\end{align}
In addition to an exponential fall-off factor, the energy $E$ dependence of
Eq.(\ref{main}) contains a pre-power factor in the high energy limit. To
obtain the linear relations for the amplitudes at each fixed mass level, we
rewrite Eq.(\ref{main}) in the following form
\begin{equation}
\frac{\mathcal{T}^{T^{n-2q}L^{q},T^{n-2q^{\prime}}L^{q^{\prime}}}}%
{\mathcal{T}^{T^{n},T^{n}}}=\left(  -\frac{2}{M\cos^{2}\phi}\right)
^{q+q^{\prime}}. \label{final}%
\end{equation}
One first notes that Eq.(\ref{final}) does not contradict with Eq.(\ref{5}),
which predict the ratios $\left(  -\frac{1}{2M}\right)  ^{q+q^{\prime}}$. This
is because for the absorption process we are considering, there is only one
kinematic variable and the usual Ward identity calculations do not apply. To
compare Eq.(\ref{final}) with the "ratios" of the Domain-wall scattering
\cite{Wall}%
\begin{equation}
\frac{\mathcal{T}^{T^{n-2q}L^{q},T^{n-2q^{\prime}}L^{q^{\prime}}}}%
{\mathcal{T}^{T^{n},T^{n}}}\mid_{Domain}=\left(  \frac{E\sin\phi}%
{M\sqrt{\left\vert M_{1}^{2}-2M^{2}-1\right\vert }\cos^{2}\phi}\right)
^{q+q^{\prime}},
\end{equation}
one sees that, in addition to the incident angle $\phi$, there is an energy
dependent power factor within the bracket of $q+q^{\prime}$. Thus there is no
linear relations for the Domain-wall scatterings. On the contrary,
Eq.(\ref{final}) gives the linear relations (of the kinematic variable $E$)
and ratios among the high energy amplitudes corresponding to absorption of
different closed string states for each fixed mass level $n$ by D-brane. Note
that since the scattering angle $\theta$ is fixed by the incident angle $\phi
$, $\phi$ is not a dynamical variable in the usual sense. Another way to see
this is through the relation of $s$ and $t$ in Eq.(\ref{st}). We will call
such an angle a \textit{geometrical parameter} in contrast to the usual
dynamical variable. This kind of geometrical parameter shows up in closed
string state scattered from generic D$p$-brane (except D-instanton and
D-particle) \cite{Dscatt,Wall}. This is because one has only two dynamical
variables for the scatterings, but needs more than two variables to set up the
kinematic due to the relative geometry between the D-brane and the scattering
plane at high energies. We emphasize that our result in Eq.(\ref{final}) is
consistent with the coexistence \cite{Wall} of the linear relations and
exponential fall-off behavior of high energy string/D-brane amplitudes. That
is, linear relations of the amplitudes are responsible for the softer,
exponential fall-off high-energy string/D-brane scatterings than the power-law
field theory scatterings.

Acknowledgments: This work is supported in part by the National Science
Council and National Center for Theoretical Sciences, Taiwan, R.O.C.

\end{document}